\documentclass[aps,prl,twocolumn,showpacs,superscriptaddress,floatfix]{revtex4-1}

\usepackage{natbib}
\usepackage{graphicx}
\usepackage{dcolumn}
\usepackage{bm}
\usepackage{amssymb}
\usepackage{microtype}
\usepackage{xfrac}
\usepackage{gensymb}
\usepackage{array}
\begin{document}
\title{Magnetic Ordering in the Ising Antiferromagnetic Pyrochlore Nd$_2$ScNbO$_7$}

\author{C.~Mauws}
\affiliation{Department of Chemistry, University of Manitoba, Winnipeg R3T 2N2, Canada}
\affiliation{Department of Chemistry, University of Winnipeg, Winnipeg R3B 2E9, Canada}

\author{N.~Hiebert}
\affiliation{Department of Chemistry, University of Winnipeg, Winnipeg R3B 2E9, Canada}

\author{M. L.~Rutherford}
\affiliation{Department of Chemistry, University of Winnipeg, Winnipeg R3B 2E9, Canada}

\author{H.~D. Zhou}
\affiliation{Department of Physics and Astronomy, University of Tennessee-Knoxville, Knoxville 37996-1220, United States}
\affiliation{National High Magnetic Field Laboratory, Florida State University, Tallahassee 32306-4005, United States}

\author{Q.~Huang}
\affiliation{Department of Physics and Astronomy, University of Tennessee-Knoxville, Knoxville 37996-1220, United States}

\author{M.~B.~Stone}
\affiliation{Neutron Scattering Division, Oak Ridge National Laboratory, Oak Ridge, Tennessee 37831, USA}

\author{N.~P.~Butch}
\affiliation{Centre for Neutron Research, National Institute of Standards and Technology, 100 Bureau Drive, MS 6100, Gaithersburg, Maryland 20899, USA}

\author{Y.~Su}
\affiliation{J\"ulich Centre for Neutron Science (JCNS) at Heinz Maier-Leibnitz Zentrum (MLZ), Forschungszentrum J\"ulich, Lichtenbergstrasse 1, 85747 Garching, Germany}

\author{E.~S.~Choi}
\affiliation{National High Magnetic Field Laboratory, Florida State University, Tallahassee 32306-4005, United States}

\author{Z.~Yamani}
\affiliation {Canadian Neutron Beam Centre, National Research Council of Canada, Chalk River, K0J 1P0, Canada}

\author{C.~R.~Wiebe}
\affiliation{Department of Chemistry, University of Manitoba, Winnipeg R3T 2N2, Canada}
\affiliation{Department of Chemistry, University of Winnipeg, Winnipeg R3B 2E9, Canada}
\affiliation{Department of Physics and Astronomy, McMaster University, Hamilton L8S 4M1, Canada}
\affiliation{Centre for Science at Extreme Conditions, University of Edinburgh, Edinburgh EH9 3JZ, UK}

\date{\today}

\begin{abstract}
The question of structural disorder and its effects on magnetism is relevant to a number of spin liquid candidate materials.  Although commonly thought of as a route to spin glass behaviour, here we describe a system in which the structural disorder results in long-range antiferromagnetic order due to local symmetry breaking. Nd$_2$ScNbO$_7$ is shown to have a dispersionless gapped excitation observed in other neodymium pyrochlores below T$_N$ = 0.37 K through polarized and inelastic neutron scattering. However the dispersing spin waves are not observed. This excited mode is shown to occur in only 14(2) \% of the neodymium ions through spectroscopy and is consistent with total scattering measurements as well as the magnitude of the dynamic moment 0.26(2) $\mu_B$. The remaining magnetic species order completely into the all-in all-out Ising antiferromagnetic structure. This can be seen as a result of local symmetry breaking due disordered Sc$^{+3}$ and Nb$^{+5}$ ions about the A-site. From this work, it has been established that B-site disorder restores the dipole-like behaviour of the Nd$^{+3}$ ions compared to the Nd$_2$B$_2$O$_7$ parent series. 


\end{abstract}
\maketitle

\section{Introduction}
Magnetic disorder induced by structural disorder has been an increasingly important issue in the search for quantum spin liquids.    One of the open questions in this field concerns the role that chemical disorder plays in the low temperature magnetic properties of frustrated magnets which can lead to effects that mimic spin liquid behaviour.   Naively, one would expect that adding chemical disorder would lead to spin glass behaviour, but even this issue is still controversial \cite{mydosh2014spin}.    Over the last few years, there have been a number of key studies of chemical disorder in quantum spin liquid candidates such as the kagome based Herbertsmithite \cite{freedman2010site, rozenberg2008disorder}, the triangular lattice YbMgGaO$_4$ \cite{paddison2017continuous, zhu2017disorder} and a variety of pyrochlore materials with tetrahedral magnetic sublattices \cite{baroudi2015symmetry, gaudet2015neutron, trump2018universal, sibille2018experimental, kadowaki2019spin}.   In the latter case, the pyrochlores with mixed nonmagnetic sites are beginning to see a significant amount of investigation, including the A$^{+1}$A'$^{+2}$B$_{2}$F$_{7}$ transition metal pyrochlores \cite{sanders2016nasrmn2f7} (where A is an alkali metal, A' an alkali earth metal and B is a magnetic transition metal), and the Ln$_{2}$B$^{+3}$B'$^{+5}$O$_{7}$ rare earth pyrochlores \cite{zouari2008synthesis, strobel2010} (where B is a transition metal and Ln is a Lanthanide).   One of the crucial issues which has not been resolved has been the effect of non-magnetic cation disorder upon the magnetic cations in tetrahedral lattices, which has broad impact to a wide variety of exotic ground states such as the quantum spin liquid and quantum spin ice.

The role of chemical disorder in dipole-octupole systems such as Nd$_{2}$Zr$_{2}$O$_{7}$ \cite{petit_observation_2016} and Nd$_{2}$Hf$_{2}$O$_{7}$ \cite{anand2017muon}  is particularly topical as chemical disorder has been widely studied in these pyrochlores\cite{blanchard2012does, ubic2008oxide}.   The chemical robustness of this unusual low temperature state, in which a large portion of the total moment remains in a gapped dynamic spin ice-like state,\cite{benton2016quantum} has never been studied until now.  This is important not only in the identification of monopole crystallization, which may be nucleated by magnetic defects \cite{sala2014vacancy, ladak2011direct}, but also in unravelling the spectroscopic signature of moment fragmentation - the coexistence of a ``divergence-full'' ordering in the elastic portion of the spectrum and a dispersing spin wave spectrum, alongside a ``divergence-free'' part \cite{brooks2014magnetic, lhotel2020fragmentation} which, in the case of Nd pyrochlore systems, remains fluctuating with a dispersionless gap in the spectrum. It is worth noting that moment fragmentation is a broad phenomenon that manifests differently and must be discussed differently in various systems  \cite{lefranccois2017fragmentation, paddison2016emergent, canals2016fragmentation}; the discussion here is limited to the case of dipole-octupolar pyrochlores. To address the issue of chemical disorder on the moment fragmentation-like ground state, we have prepared the mixed B-site pyrochlore Nd$_{2}$ScNbO$_{7}$.   In this system, the rare earth Nd$^{+3}$ fully occupies the 16c site but the 16d site contains a solid solution of Sc$^{+3}$ and Nb$^{+5}$ cations, with a N\'eel temperature of 0.37 K (Fig~\ref{fig1}).  Remarkably, we find that even in the limit of severe chemical disorder on the B-site that the system shows key experimental signatures of moment fragmentation, with coexisting magnetic Bragg peaks and diffuse scattering observed beneath T$_N$.    Therefore, in the case of the Kramers doublet systems which allow the dipolar and octupolar components of the moment to transform independently, we show here that chemical disorder does not play a strong role in the underlying physics at low temperatures.  Instead the maintenance of symmetry is required for retaining an octupolar behaviour, and the destruction of that symmetry, even by next-nearest neighbour chemical disorder, restores dipole-like behaviour compared to the parent Nd$_2$B$_2$O$_7$ series.   

\section{Methods}

Polycrystalline Nd$_2$ScNbO$_7$ was prepared by conventional solid state synthesis at 1400~$^{o}$C. The single crystal used in this study was prepared by the floating zone method under flowing air at a growth rate of 8 mm/hr.  The quality of the single crystal was confirmed by x-ray Laue diffraction and 4-circle neutron diffraction at the Chalk River Laboratories. The sample is largely single crystalline barring a few grains ($<$1 \% by intensity) in comparison to the dominant phase's Bragg peak intensities. The pyrochlore structure exists as a supercell of the parent defect fluorite, where the cations form two distinct sublattices and the oxygens in the pyrochlore preferentially occupy the 8b Wycoff position over the 8a position. This gives insight into the common defects in pyrochlore systems, where cations may randomly distribute themselves over the sublattices and oxygens may occupy the 8a position. Powder x-ray diffraction on a crushed portion of the single crystal was refined, and all cations were allowed to refine to either cationic site (the Rietveld refinement yielded a R$_{wp}$ of 11.2 with $<$ 0.2 \% site mixing of the cations).  Another common defect among pyrochlores is on the oxygen site, with the oxygens from the fully occupied 8b site moving to the systematically vacant 8a site.  Neutron total scattering results did not require any 8a oxygen defects to adequately describe the results. 

\begin{figure}[t]
\linespread{1}
\par
\includegraphics[width=3.3in]{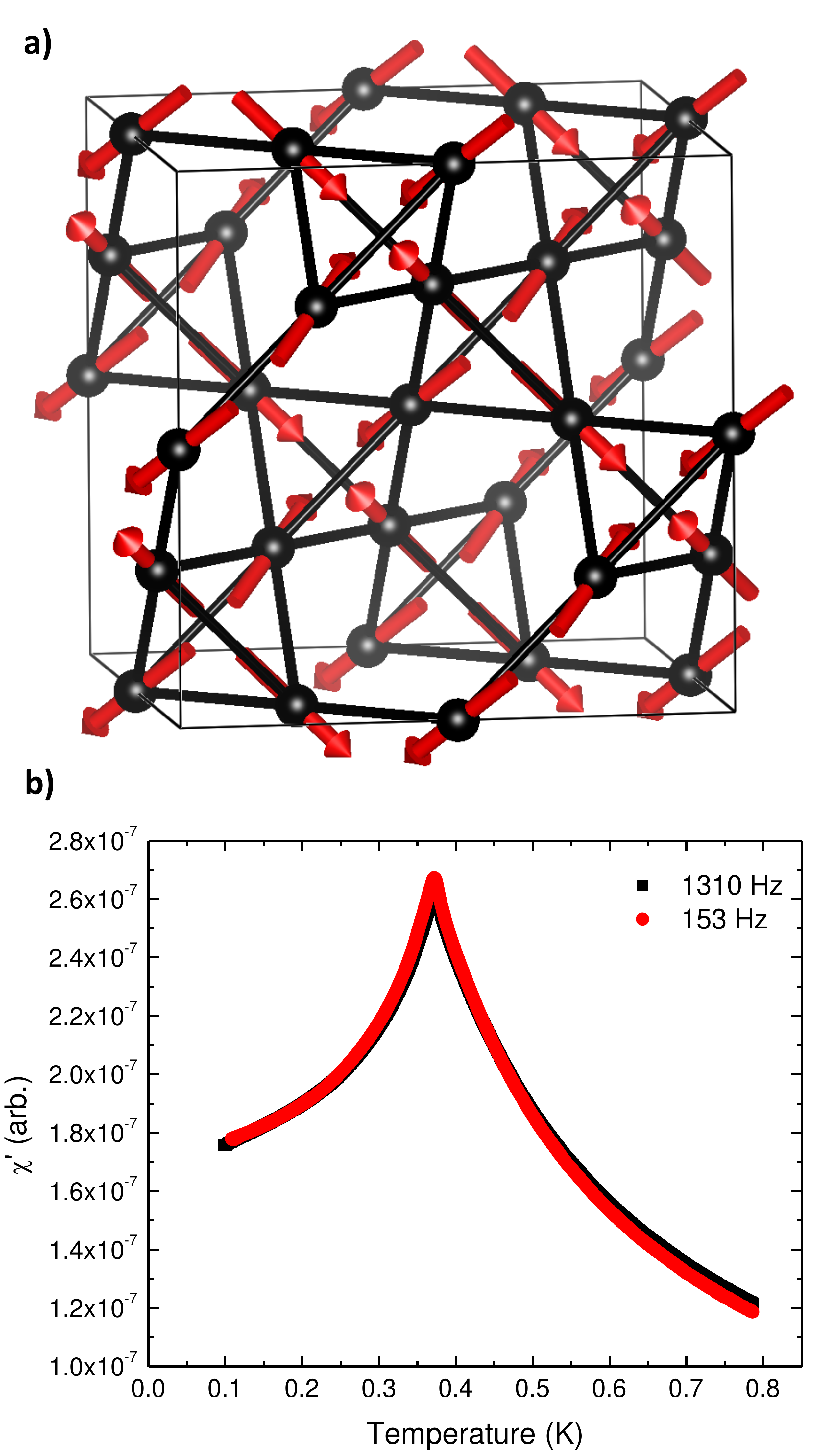}
\par
\caption{(a) The pyrochlore lattice with antiferromagnetic all-in, all-out (AIAO) order along a local $<$111$>$ Ising axis. (b) AC susceptibility of powder Nd$_2$ScNbO$_7$ noting a peak near T$_N$ = 0.37 K. No frequency dependent shift is observed.}
\label{fig1}
\end{figure}

AC susceptibility was performed at the National High Field Magnet Lab using an in-house SQUID mounted in a dilution refigerator cryostat, at the frequencies shown in figure \ref{fig1}. The 1310 Hz data in figure \ref{fig1} was shifted vertically by 0.32 x 10$^{-7}$ cm$^{-3}$ to account for an issue with the AC phase during measurement, in order to make the data more readable as total intensity was not relevant to our results. No frequency shift was detected for the susceptibility as a function of temperature within this range. DC magnetization measurements were taken on a Quantum Design PPMS with a base temperature of 1.8 K, using a 6.1 mg powder sample. 

Inelastic neutron scattering measurements were taken on a powder sample using the SEQUOIA spectrometer at Oak Ridge National Laboratories \cite{granroth2010sequoia} with multiple incident energies (E$_i$ = 10.5, 60 and 140 meV) at temperatures of 5, 100, and 250 K. The majority of the crystal field spectral weight was found to be distributed continuously from 10-50 meV. To accurately remove the phonon interference we subtract off a spectrum of the La$_{2}$ScNbO$_{7}$ phonon standard, normalized by calculated Bragg peak intensities to a self shielding factor of 0.82. The remaining scattering is solely due to crystal field excitations, to a good approximation. Additionally the higher energy excitation present at 106.5(1) meV is observed, and remains very broad at 14.1(2) meV FWHM.  At an incident energy of 140 meV the sharp excitations cannot be resolved from the broad features, and the lower broad excitation is not resolved from the elastic line, which is why the relative intensity of the 106.5 meV peak is not fit in the data.   Additional inelastic neutron scattering measurements were performed using the DCS at NIST (CHRNS) on a crystal aligned in the (HHL) plane.   A dilution refigerator was used to reach low temperatures with the sample fixed with copper wire for thermal conductivity.   In order to improve the low intensity inelastic signal, measurements were averaged over Q.  A 20 K empty cryostat run was used as a background.

X-ray pair distribution function analysis (PDF) was performed at the Canadian Light Source using the BXDS-WHE instrument.  A wavelength of 0.155 \AA~ was used to provide a Q$_{max}$ of 28 \AA$^{-1}$ at a temperature of 90 K (to minimize atomic vibrations present at room temperature).   Data were reduced and the diffraction patterns fit in GSAS II, and the PDF data were fit utilizing PDFgui. \cite{farrow2007pdffit2, toby2013gsas}.   Calibrations were performed using LaB$_6$ and Ni standards.   Neutron PDF was performed at NOMAD (Oak Ridge National Laboratory).   Data were taken in a quartz ampule using a 0.539 g sample at 300 K.   Vanadium standards were used for absolute intensity calibration.   Data were reduced using in house software at ORNL.   

Polarized neutron scattering measurements were performed on a single crystal of Nd$_2$ScNbO$_7$ using the DNS spectrometer at FRM-II.    The crystal was aligned in the (HHL) plane for the measurements using a wavelength of 4.1916 \AA.   Measurements were taken in a dilution refrigerator from 100 mK to 1300 mK, and then to higher temperatures of 5 K and 10 K.

\section{Results}
\subsection{Crystal Electric Field and Magnetization}

\begin{figure}[t]
\linespread{1}
\par
\includegraphics[width=3.3in]{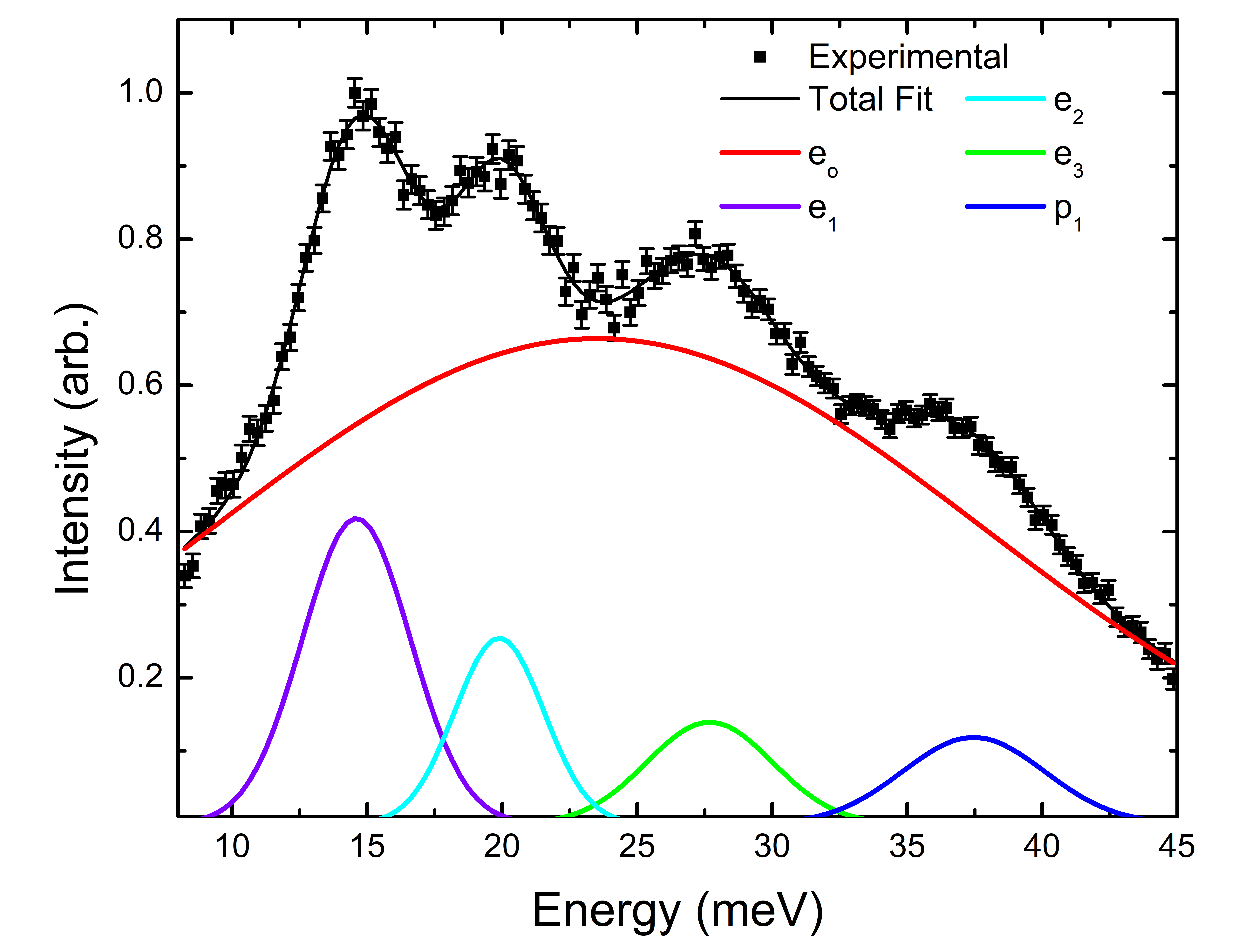}
\par
\caption{Electronic excitation spectrum at 5 K of Nd$_2$ScNbO$_7$ taken on SEQUOIA using La$_2$ScNbO$_7$ at 5 K as a phonon correction, and integrated over Q from 2.5-4 \AA$^{-1}$. Discrete crystal electric fields are labeled e$_{i}$ with the broad anomaly labeled e$_{0}$, and p$_{1}$ is an imperfectly subtracted phonon.}
\label{fig2}
\end{figure}

\begin{figure}[t]
\linespread{1}
\par
\includegraphics[width=3.3in]{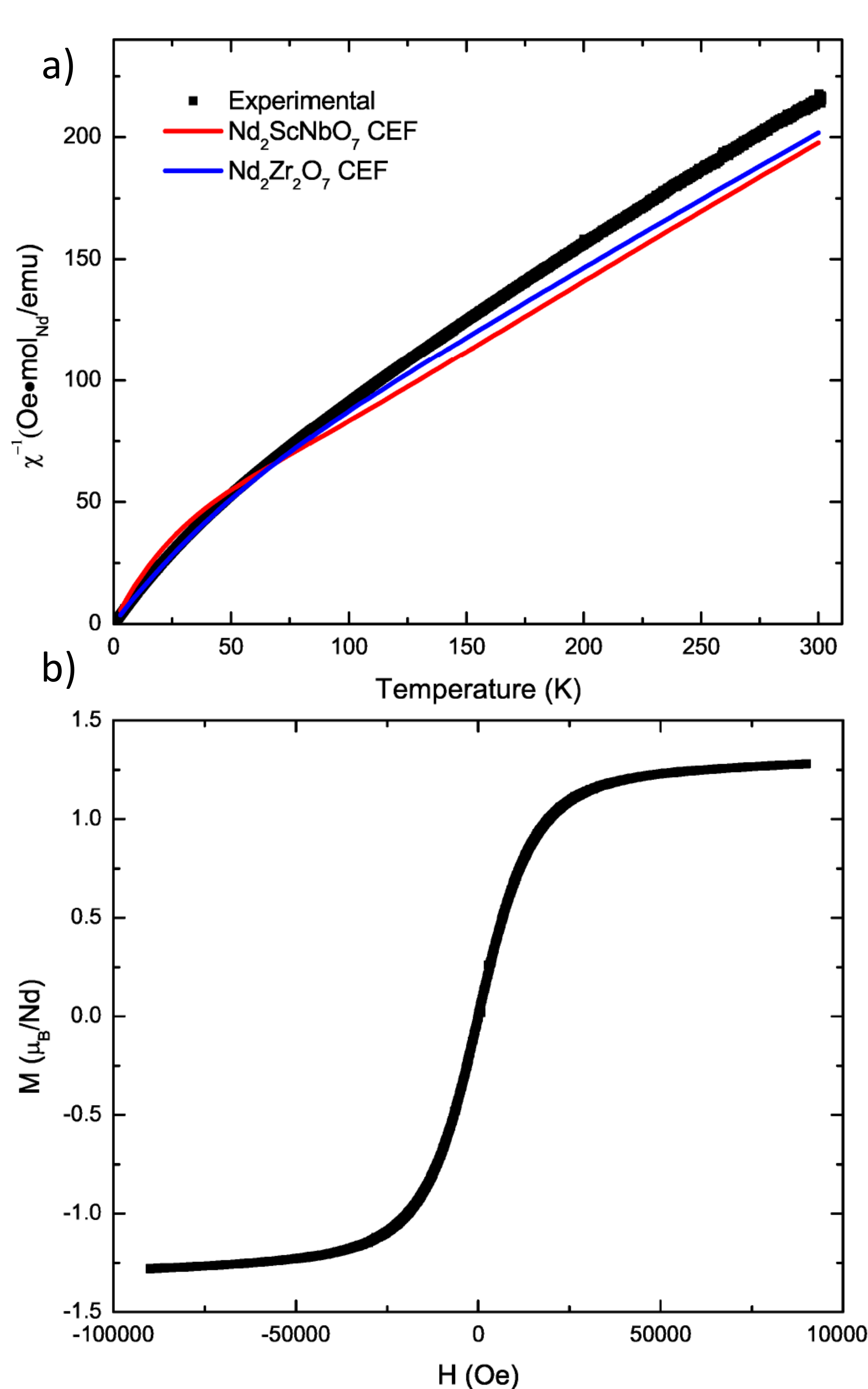}
\caption{(a) Inverse DC magnetic susceptibility in a 1000 Oe applied field on a powder sample of Nd$_2$ScNbO$_7$ (Black), with calculated susceptibilities for the experimentally determined crystal field (Blue) and calculated values for Nd$_2$Zr$_2$O$_7$ (Red), taken from the crystal field parameters presented in X. Zu\emph{~et.~al.} \cite{xu_magnetic_2015}. (b) Magnetization measurements on powder Nd$_2$ScNbO$_7$ at 1.8 K with a maximum applied field of 90000 Oe.}
\label{magfig}
\end{figure}

The initial effects of extreme chemical disorder on the B-site in pyrochlores can readily be seen in the broadening of crystal field excitations.   Neutron crystal electric field (CEF) spectroscopy was performed on Nd$_2$ScNbO$_7$ to explore these effects and characterize the local rare earth environment. To remove phonon interference the non-magnetic structural analog La$_2$ScNbO$_7$ was used as a background subtraction (Fig~\ref{fig2}).   The crystal field peaks are not sharp in energy, which is a natural consequence of the heterogeneous local distortions imposed by the charge disorder. A broad inelastic excitation centred around 25 meV energy transfer follows the Nd$^{+3}$ magnetic form factor (Appendix) and is assumed to be the result of the three very broad excited crystal field doublets overlapping within the region (Fig~\ref{fig2}). Another well-developed crystal field doublet appears at 108 meV energy transfer (Appendix). On top of this large, broad crystal field excitation, four discrete excitations are observed. The excitation labeled p$_1$ is an optical phonon visible in La$_2$ScNbO$_7$ that is not perfectly subtracted (Appendix).  The remainder of the excitations indexed as e$_i$ are designated as crystal fields.  The four observable CEF excitations are fit to the J = 9/2 ground state manifold for Nd under an assumed hexagonal local symmetry (D$_{3d}$ or D$_d$) \cite{walter1984treating} which require six crystal field parameters. The crystal field scheme was fit using the Spectre \cite{boothroydspectre} crystal field analysis package. Using the Wybourne description of the crystal field for a hexagonally symmetric system (D$_3$ or D$_{3d}$):

\begin{equation}
H_{CEF} = B_{2}^{0} C_{2}^{0}+B_{4}^{0} C_{4}^{0}+B_{6}^{0} C_{6}^{0}+B_{4}^{3} C_{4}^{3}+B_{6}^{3} C_{6}^{3}+B_{6}^{6} C_{6}^{6}
\label{eq: 1}
\end{equation}

the set of parameters were refined to  $B_{2}^0 = -36.36 ~$meV$, B_{4}^0 = 442.4 ~$meV$, B_{6}^0 = 186.2 ~$meV$, B_{4}^3 = 173.7 ~$meV$, B_{6}^3 = -66.75 ~$meV$, $ and $ B_{6}^6 = 118.8 ~$meV. This results in the crystal field scheme presented in Table I, which gives a completely Ising moment of 2.0 $\mu_{B}$. This idealized crystal field scheme does not account for the large broad crystal field excitation labeled e$_0$ (Fig~\ref{fig2}). Comparing the spectral weight of the broad and discrete excitations shows that the discrete crystal fields only comprises 14(2) $\%$ of the spectral weight. The calculated magnetic susceptibility from these parameters can be compared against experiment (Fig \ref{magfig}, a). Both the fitted crystal field scheme and the crystal field scheme for Nd$_2$Zr$_2$O$_7$ \cite{xu_magnetic_2015}, which do vary from each other significantly, are insufficient in explaining the bulk magnetic susceptibility of Nd$_2$ScNbO$_7$.  This calculation does not agree with the bulk susceptibility, as this crystal field fit only represents 14(2) $\%$ of the Nd$^{+3}$ ions. For completion the calculated susceptibility for Nd$_2$Zr$_2$O$_7$ is also shown disagreeing with the experimental data, although arguably fitting slightly better than the refined CEF. As the majority of ions contain distinct local fields, it is unlikely that any single crystal field scheme can meaningfully describe the bulk data. Because of this, the total moment should be considered tentative. Another way of attempting to extract the total moment is through magnetic saturation, (Fig~\ref{magfig}, b) via DC Magnetometry. For a purely Ising system, the magnetization saturates at 1/2~$\mu_{total}$, which gives a total moment of 2.5(1) $\mu_{B}$. However this Ising assumption should only be considered an upper limit as there is significant overlap of the broad excited crystal fields with the ground state which would allow other anisotropies, resulting in a saturation moment higher than 1/2~$\mu_{total}$, as the system approaches a Heisenberg anisotropy. This gives us a reasonable range of 2.0-2.5~$\mu_B$ consistent with other Nd pyrochlores. 

\begin{table}
\caption{
\label{tab: 1} 
Crystal electric field results for  Nd$_2$ScNbO$_7$ within the ground state J-manifold. Unlabelled intensities are due to being unable to extract relative intensities while using a higher E$_i$. Basis with contributions less than 0.01 are excluded, including higher multiplet contributions.}
\begin{ruledtabular}
\begin{tabular}{cc|cc|cccccc}
E$_{obs}$(meV)&I$_{obs}$&E$_{fit}$(meV)&I$_{fit}$&$\pm\frac{1}{2}$&$\pm\frac{3}{2}$&$\pm\frac{5}{2}$&$\pm\frac{7}{2}$&$\pm\frac{9}{2}$\\ \hline
0.0& &0&&0&0.51&0&0&0.85\\
14.6(1)&1&13.52&1&0.77&0&0.63&0.01&0\\
19.9(1)&0.5(1)&20.92&0.46&0&0.85&0&0&0.50\\
27.7(1)&0.4(2)&28.23&0.5&0.55&0&0.66&0.47&0\\
106.5(1)& - &104.86& - &0.30&0&0.38&0.86&0\\
\end{tabular}
\end{ruledtabular}
\end{table}

\begin{figure}[t]
\linespread{1}
\par
\includegraphics[width=3.3in]{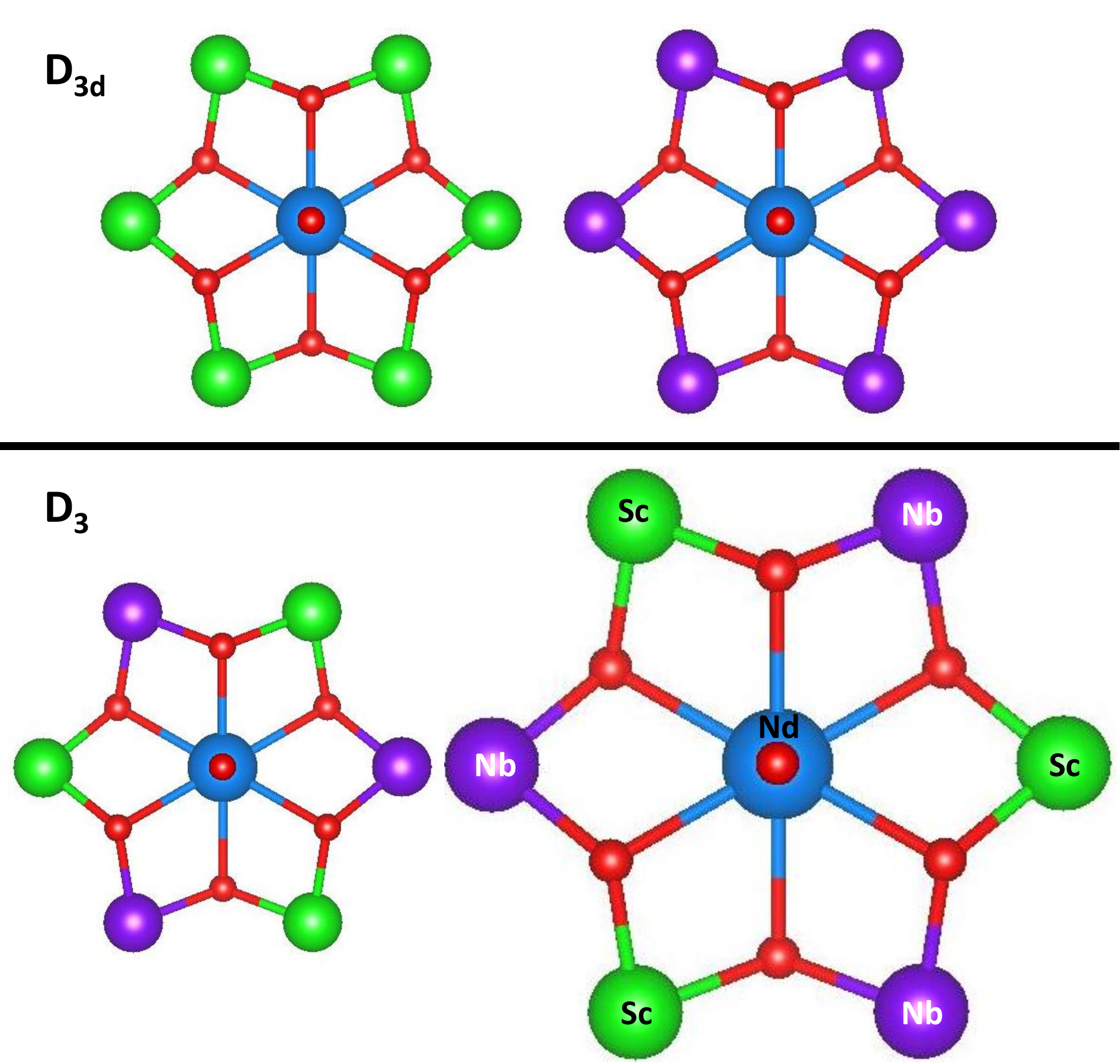}
\par
\caption{Possible local environments of Sc$^{+3}$ and Nb$^{+5}$ that retain a C$_3$ rotation centre, of the possible 2$^6$ configurations. The D$_{3d}$ configurations maintain the crystallographic site symmetry, whereas the D$_3$ configurations have lost inversion. The configuration in the bottom right is enlarged to accommodate labelling that applies to all configurations.}
\label{fig3}
\end{figure}

\begin{figure}[t]
\linespread{1}
\par
\includegraphics[width=3.3in]{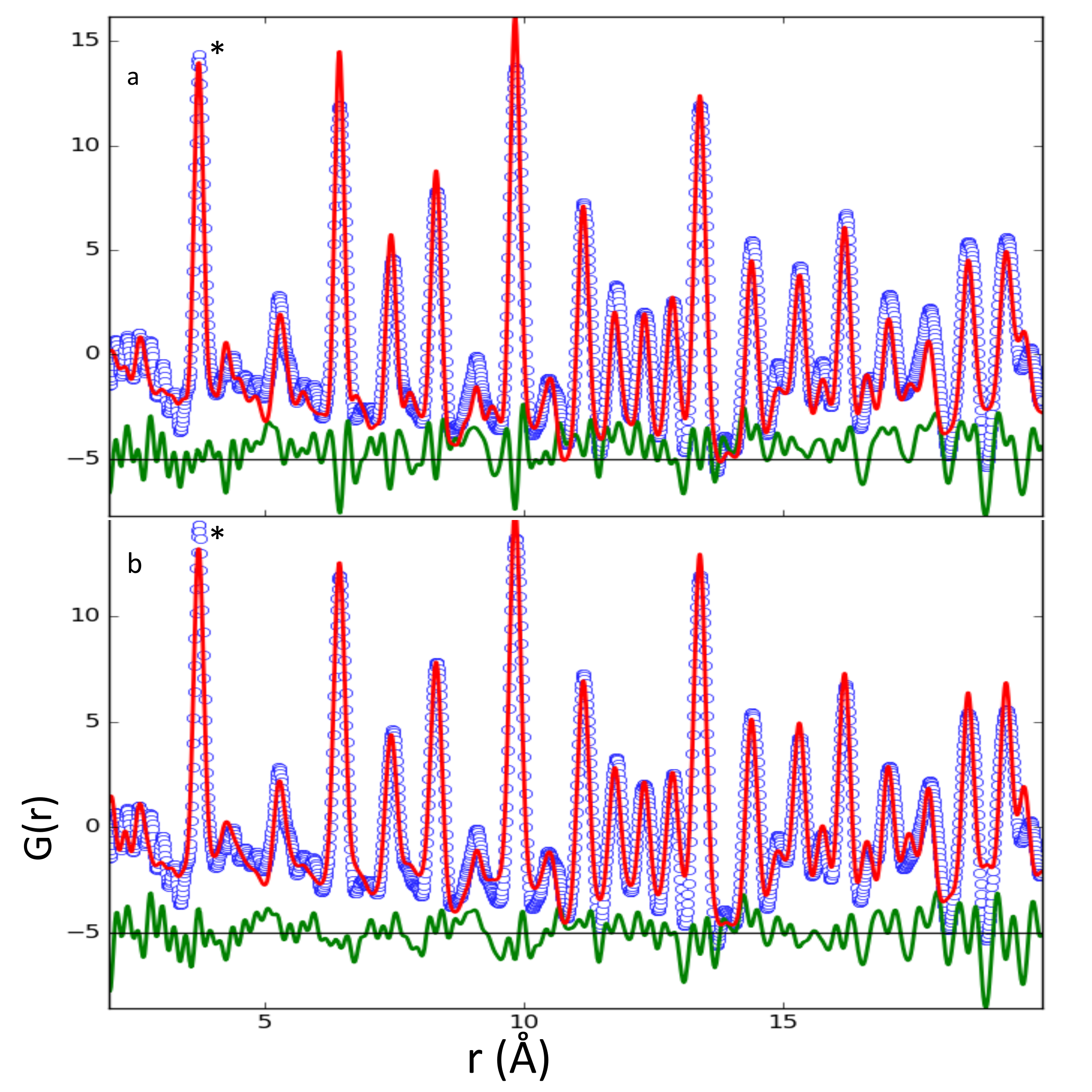}
\par
\caption{X-ray PDF data obtained on a ground single crystal of Nd$_2$ScNbO$_7$ at 100 K. Blue data points are experimental, the red line shows the respective fits, and the green line shows the difference. Fit (a) shows the fit described in text that quantifies the relative ratio of B-site correlations. Fit (b) is a fit to the data using a conventional unit cell approach. The peak denoted * contains the nearest neighbour Nd-Nd, Nd-Sc, Nd-Nb, Sc-Sc, Sc-Nb, and Nb-Nb correlations, all with crystallographically equivalent distances.  }
\label{xraypdf}
\end{figure}

\begin{figure}[t]
\linespread{1}
\par
\includegraphics[width=3.3in]{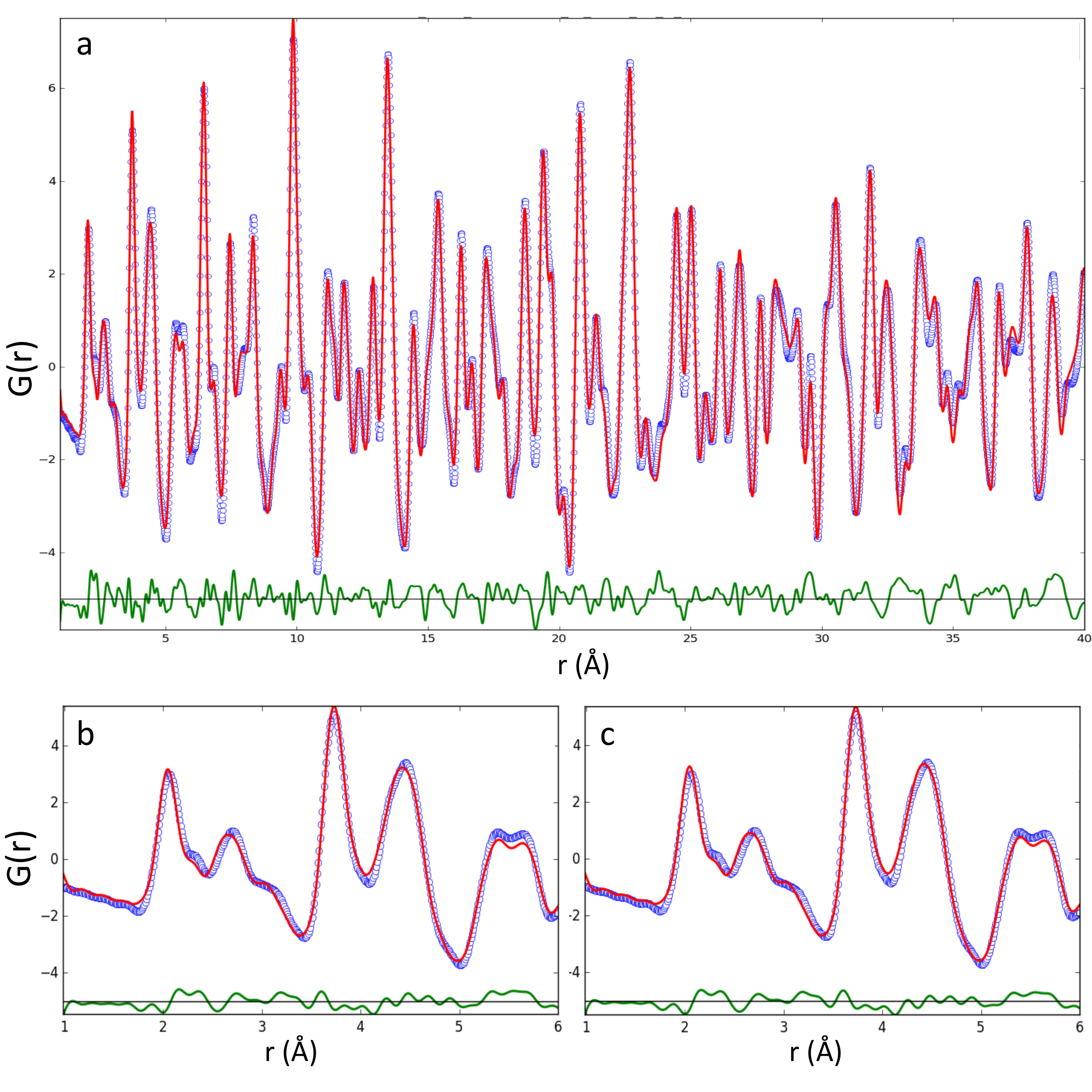}
\par
\caption{Neutron PDF data obtained on a ground single crystal of  Nd$_2$ScNbO$_7$ at 100 K.  Blue data points are experimental, the red line shows the respective fits, and the green line shows the difference. Panel (a) shows a full fit of the primitive crystallographic lattice with random Sc-Nb correlations. Panel (b) shows the a fit to 6 \AA~ using the parameters from (a) and fitting only the scale factor. Panel (c) shows a fit to 6 \AA~ using the results of (a) and also fitting the Sc-Nb, Nb-Nb and Sc-Sc correlations.}
\label{fig5}
\end{figure}

The coexistence of broad and sharp crystal field excitations may be explained by a distribution of local environments about the neodymium site.   Our initial assumption is that the discrete peaks originate from the most probable local coordinations, causing them to stand out against the background of varying excitations. If these discrete excitations originated from the $\overline{D_{3d}}$ double group of the D$_{3d}$ retaining configurations, which exists for two possible arrangements of Sc$^{+3}$ and Nb$^{+5}$ next nearest neighbour configurations (Fig~\ref{fig3}), in which all Sc or all Nb form a hexagon about the A-site, we could assume a statistical distribution of Sc and Nb on the B-site and would expect 1/32 (3$\%$) of the magnetic ions to retain $\overline{D_{3d}}$ symmetry.  However these two configurations would likely split the CEF excitations within our energy resolution, and this is not observed.  Additionally, grouping of the Sc and Nb ions seems unlikely due to the large charge repulsion that would occur from clusters of Nb$^{5+}$ cations.  If, for example, we assume that each tetrahedron has an average +4 charge (two Sc$^{+3}$ and two Nb$^{+5}$) only 0.4 $\%$ of magnetic ions should retain $\overline{D_{3d}}$ symmetry.  Instead what is likely being observed as discrete excitations in the CEF spectrum are the $\overline{D_{3}}$ systems (Fig~\ref{fig3}).  Under $\overline{D_{3}}$ symmetry equation (1) still holds and we should still expect a similar crystal field spectrum to other Nd pyrochlores. The fact that the discrete e$_{i}$ excitations stand out against the background of Nd ions with various local environments suggests that these discrete excitations originate from the most probable local configuration of ions about Nd$^{+3}$. 

The $\overline{D_{3}}$ configurations appearing well above the statistical expectation of a fully disordered system is due to the charge repulsion of Sc and Nb ions, as this configuration will provide the least charge repulsion, or potentially the least local strain due to ionic size discrepancies. This demonstrates that there are strong short-range correlations between Sc and Nb ions. If we assume the charge ice rule of two Sc$^{+3}$ and two Nb$^{+5}$ per B-site tetrahedron, giving each tetrahedron a net +4 charge per ion, 17.6 $\%$ of our Nd environments should have the D$_3$ configuration. With 14(2) \% of the Nd ions containing this local environment, it suggests that the system does not have perfect charge ice behaviour but strong correlations approaching charge ice. 

\subsection{Pair Distribution Function Analysis}

In order to explore the possibility of charge ice ordering, and understand the nature of the Sc-Nb correlations, x-ray and neutron pair distribution function (PDF) analyses were performed. Despite the high resolution (roughly 0.2 \AA~  full width at half maximum (FWHM)) of the x-ray data, there was no splitting of peaks noted by an asterix (*) in figure~\ref{xraypdf}, which would contain information about the A-A, B-B and A-B  nearest neighbour correlations.    The fits to the data in Figure~\ref{xraypdf} are shown using two methods.    Fit (b) utilized the crystallographic fit to the Bragg peaks and parameterizes several peak shape variables as well as scale and unit cell dimensions, yielding a scale (0.63(1)), a unit cell (10.478(2) \AA), a ~$\frac{1}{r^2}$ peak sharpening $\delta_2$ (3.7(2) \AA$^2$), and an isotropic atomic displacement factor U$_{iso}$ (0.0035 \AA$^2$) over a range of 2-20 \AA.   This provides a reasonable representation of the data, but with notable errors of underweighing the peak denoted * (which has the cation correlations), and overweighing of the Sc-Nb correlation peaks beyond nearest neighbour interactions.  Fit (a) (Fig. \ref{xraypdf}) represents an attempt to better describe the cation correlations (the peak denoted * in the data).   To obtain this fit, the parameters obained from the long-range fit were fixed in order to obtain a low parameter fit.   The PDF was fit from 2.8 - 4 \AA~  fitting only a relative ratio of Sc-Nb, Nb-Nb and Sc-Sc correlations such that the sum of the correlations were fixed and the number of Nb-Nb, Sc-Sc correlations were equal, giving a relative ratio of 0.76:0.12:0.12 Sc-Nb:Sc-Sc:Nb-Nb correlations (Fig. \ref{xraypdf}, a).  This resulted in a relative error of 200 percent of the scale factor, illustrating the loss of information due to poor discernibility of relatively light species such as oxygen (Z = 8) and scandium (Z = 21) compared to heavier species such as Nb (Z = 41) and Nd (Z = 60) from x-ray scattering.  This is a common issue with x-ray PDF on high symmetry systems where cations with large electronic densities overlap with identical nearest neighbour bond lengths to ions of interest.  
\begin{figure}[t]
\linespread{1}
\par
\includegraphics[width=3.3in]{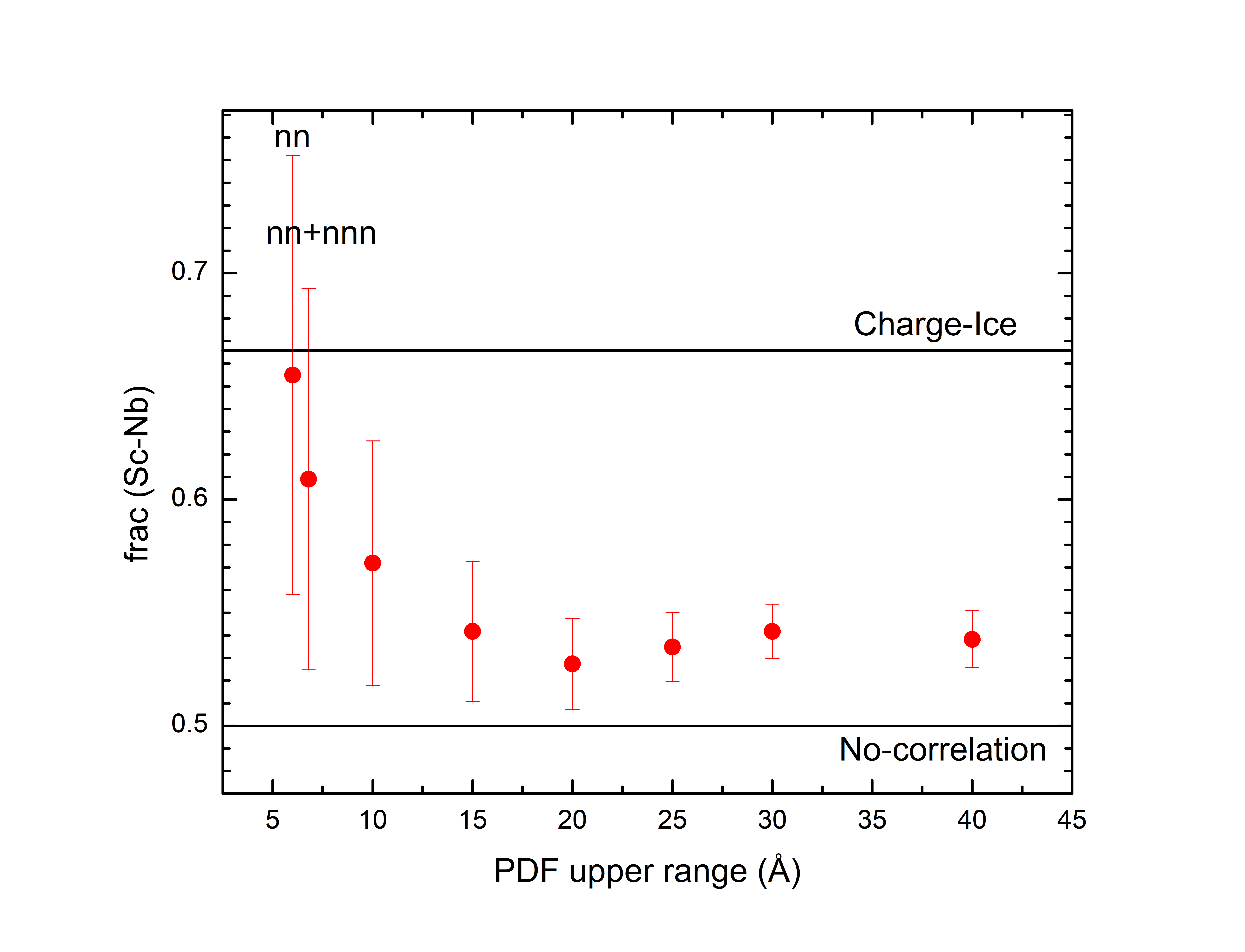}
\par
\caption{The fraction of Sc-Nb correlations, with the rest being Sc-Sc and Nb-Nb in equal proportion as a function of the fitting range maximum. The line at 2/3 represents the theoretical result of charge ice, the line at 1/2 is the expected result of a system with random occupancy, which the fit should trend to at long upper range.  The abbreviation nn represents the fit with only nearest neighbour B-B correlations, and nn+nnn represents the first two correlation spheres, with later fits including longer range correlation spheres.}
\label{fig6}
\end{figure}

\begin{figure}[t]
\linespread{1}
\par
\includegraphics[width=3.3in]{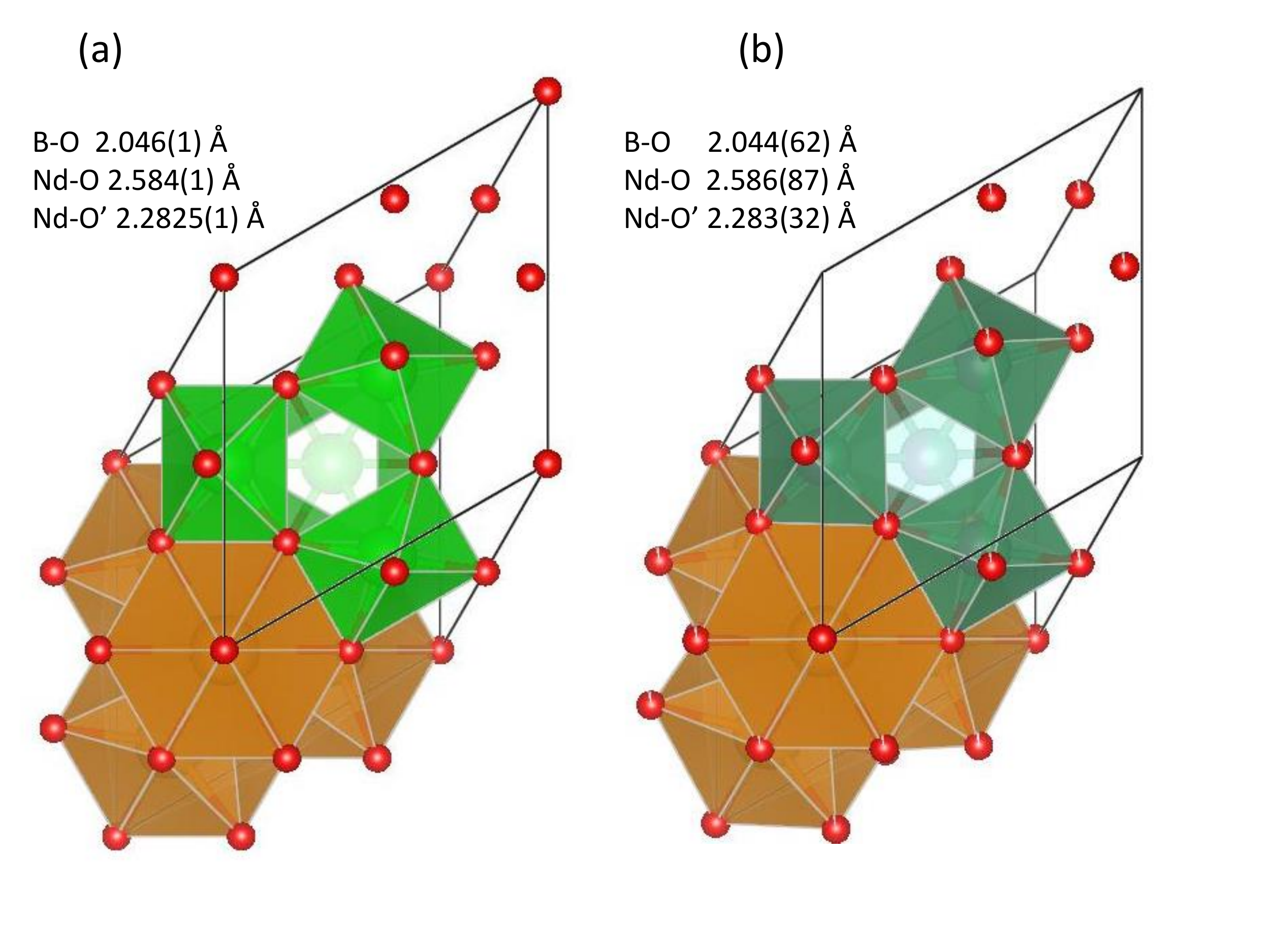}
\par
\caption{(a) Refined crystallographic Fd$\bar3$m unit cell from neutron diffraction, shown in the primitive lattice. Bond length errors are extracted from refinement errors. (b) Refined primitive unit cell from neutron PDF for Nd$_2$ScNbO$_7$, the P1 cell included free refinement of oxygen positions. Orange polyhedra contain Nd, green polyhedra contain Sc/Nb. Bond length errors are extracted as a standard deviation from the distribution of oxygen bond lengths. }
\label{fig7}
\end{figure}

While no meaningful conclusion can be drawn from the X-ray PDF with respect to B-site correlations, due to the low intensity weighings of Sc and Nb correlations relative to overlapping Nd-Nd correlations, neutron PDF can be more instructive. The increased relative intensities of the oxygen peaks allowed a better refinement of the scale factor over the short fitting window of 1-6 \AA, which was impossible with the X-ray data.   Better resolution of the cations was also obtained as there is considerable contrast between the scattering length of Sc (12.3 fm) compared to Nd (7.69 fm) and Nb (7.05 fm) \cite{sears1992neutron}.  Initial refinements of the oxygen positions in the standard Fd$\bar3$m cell did not work well.  Instead, a refinement was made from 1 - 40 \AA~ (Fig~\ref{fig5}, a) of the primitive unit cell represented in P1 symmetry, refining all of the oxygen positions freely, and assuming random B-B cation correlations, yielding a scale (0.97(1)), unit cell (7.4554(6) \AA), a ~$\frac{1}{r}$ peak sharpening $\delta_1$ (1.4(2) \AA), and U$_{iso}$ (0.0064(2) \AA$^2$). This P1 cell yielded a unit cell similar to the crystallographic cell, with very similar average bond lengths, but the larger, anisotropic, distribution better represented the data, than any refinement with the symmetrized cell could achieve (Fig.~\ref{fig7}). Using this cell, a similar fit to the neutron data was performed from 1-6 \AA~ (Fig~\ref{fig5}, b, c) with more reliable results (R$_w$ = 0.110) compared to the random B-site occupancy (R$_w$ = 0.112). Of the nearest neighbour B-B site correlations, 0.65(9) refined to Sc-Nb with the rest being equal weights of Sc-Sc and Nb-Nb, close to the expected charge ice value of 2/3 and due to the lower error bars, a randomized orientation of Sc and Nb can be ruled out.   To confirm that this is a nearest neighbour effect, the same method of fitting was used to longer ranges (Fig~\ref{fig6}), where the interactions should approach statistical values assuming this is a short range interaction.   Although this error is still too large to confirm true charge ice behaviour, it is sufficient to conclude that there are correlations that prefer Sc-Nb correlations above those expected of random occupancy.


The general ice configuration appears in many types of frustrated systems that exist on corner shared tetrahedra. Frustration from bond disorder (covalent or hydrogen bonding) leads to the ice rules within water ice with two bonding and two non-bonding hydrogens on each tetrahedron. Ising ferromagnetic interactions gives rise to the ice rule in spin ice, with two spins pointing into each tetrahedron and two out \cite{fennell2009magnetic}. Recently charge ice behaviour on a fluoride pyrochlore lattice was confirmed in CsNiCrF$_6$ \cite{fennell2019multiple}. Another charge ice system Cd(CN)$_2$ has distortions of the cyanide tetrahedra toward or away from Cd centres \cite{fairbank2012charge}. As far as we are aware Nd$_2$ScNbO$_7$ is the first case of possible charge ice behaviour within the reported oxide pyrochlores.  

\subsection{Magnetic Neutron Diffraction and Spectroscopy}

\begin{figure}[t]
\linespread{1}
\par
\includegraphics[width=3.3in]{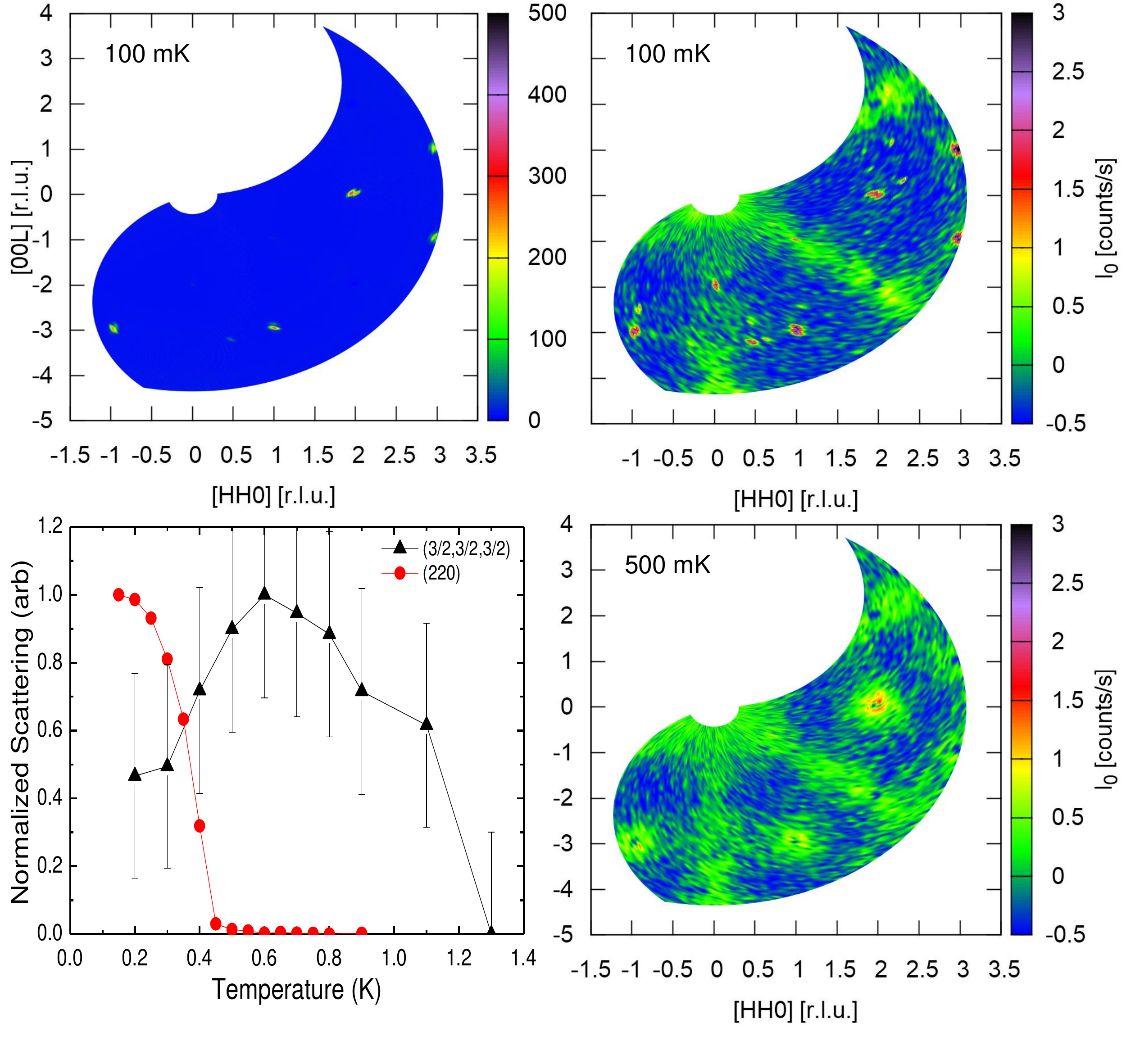}
\par
\caption{Spin Flip difference scattering with polarization in the [H,-H,0] plane shown on two intensity scales (left and right) obtained on DNS at FRMII. The temperature dependence of an antiferromagnetic Bragg peak (220) and the spin ice like diffuse scattering (3/2,3/2,3/2) (bottom left). Spin Flip difference scattering with polarization in the [H,-H,0] plane above the N\'eel order (bottom right). }
\label{fig8}
\end{figure}

To determine the effect of these correlations on the magnetic ordering of the system, polarized neutron diffraction was performed on a single crystal of Nd$_2$ScNbO$_7$.  Magnetic Bragg peaks associated with Ising antiferromagnetic order were observed along with a weak signal of diffuse scattering along the [HHH] and [00L] directions consistent with the spin ice structure factor \cite{fennell2009magnetic}. The moment associated with the Ising antiferromagnetic structure was refined from the total scattering of the two magnetic peaks (220) and (113) and four structural peaks, all taken from one quadrant [HH-L] as depicted (Fig \ref{fig8}), to minimize the effects of anisotropic scattering. SARAh representational analysis \cite{wills2000new} and FullProf Suite \cite{rodriguez1993recent} software packages were used to determine the magnetic basis states and refine the data. The refined moment shows 2.2(4) $\mu_{B}$ contributing to the long-range antiferromagnetic state, comprising the majority of the total scattering moment which lies in the range of 2.0-2.5 $\mu_B$. This shows that the antiferromagnetic ordered moment comprises much more than 0.8 $\mu_{B}$ of the total 2.3  $\mu_{B}$ moment seen in Nd$_2$Zr$_2$O$_7$\cite{lhotel_fluctuations_2015}, or the 0.6 $\mu_{B}$ of the total 2.5 $\mu_{B}$ seen in Nd$_2$Hf$_2$O$_7$\cite{anand2015observation}. Although some experiments have reported larger moments in different materials \cite{xu_magnetic_2015}, this moment of 2.2(4) $\mu_{B}$ out of a maximum 2.5 $\mu_{B}$ remains abnormally high (see Table II). 

\begin{table}
\caption{
\label{tab: 1} 
Comparison of the ordered all-in, all-out (AIAO) moment in the known Neodymium pyrochlores}
\begin{ruledtabular}
\begin{tabular}{c|c|c}
Pyrochlore&AIAO moment ($\mu_B$)&Reference\\ \hline
Nd$_2$ScNbO$_7$ &2.2(4)&This work\\
Nd$_2$Zr$_2$O$_7$&1.26(2)&J. Xu \em{et al}, 2015\\
Nd$_2$Zr$_2$O$_7$&0.8(1)&S. Petit \em{et al}, 2016\\
Nd$_2$Sn$_2$O$_7$&1.708(3)&A. Bertin \em{et al}, 2015\\
Nd$_2$Hf$_2$O$_7$&0.62(1)&V.~K.~Anand \em{et al}, 2015\\
CEF limit& 2.0-2.5 &This work \\
\end{tabular}
\end{ruledtabular}
\end{table}

Despite this large, nearly fully ordered moment, some evidence of moment fragmentation-like features are clearly seen in the coexisting spin ice like scattering beneath T$_N$ (Fig~\ref{fig8}). There exists clear diffuse scattering at low temperatures along the [HHH] and [00L] directions associated with spin ice scattering. True pinch points at (111), and (002) are not observed but this is likely due to the fact that this is an energy integrated spectrum and this scattering is not solely elastic in nature. This scattering feature is also qualitatively much too weak to be comprised of the majority of the total moment, in agreement with the large moment observed on the Bragg peaks.    

\begin{figure}[t]
\linespread{1}
\par
\includegraphics[width=3.3in]{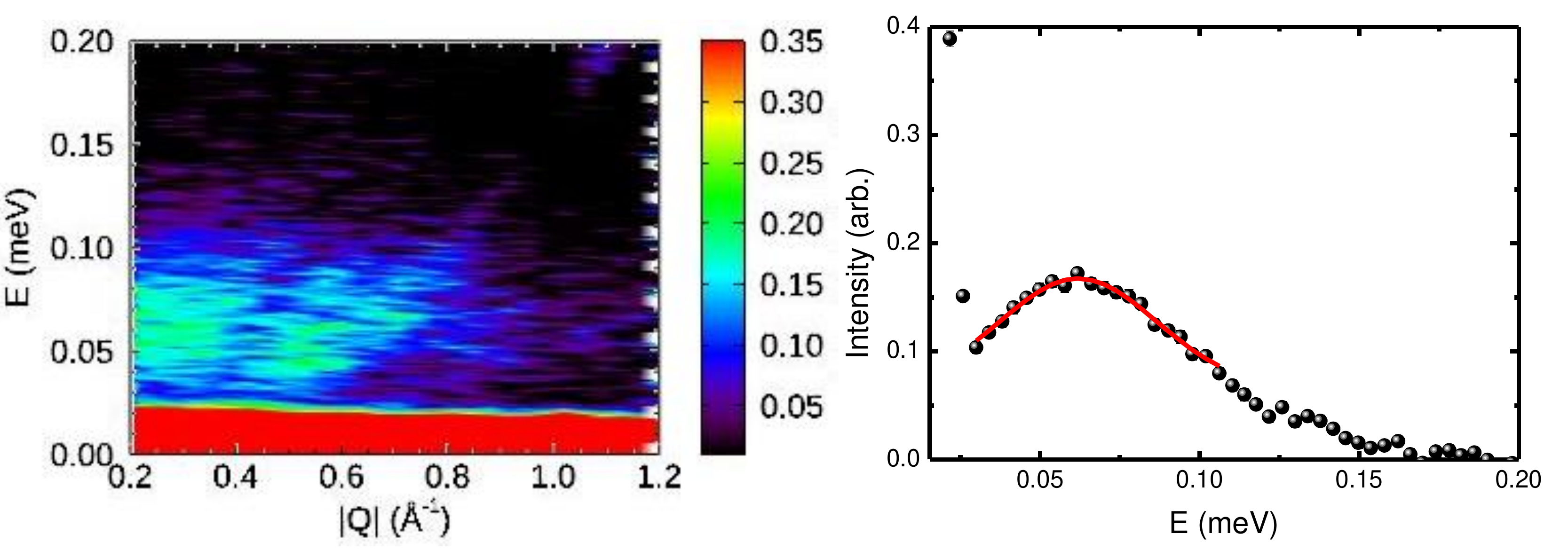}
\par
\caption{Inelastic spectroscopy using the DCS at NIST at 100 mK with an incident wavelength of 8 \AA, obtained along [H,H,H] integrating over $\pm$ 0.25 r.l.u along [-2H,H,H] (left), shown also as a cut along energy using $|Q|$ [0.2,1] (right). The red line is a guassian fit to the peak over the region shown. }
\label{fig9}
\end{figure}

To confirm the existence of moment fragmentation-like features, energy resolved measurements of the spin ice like scattering are required \cite{petit_observation_2016}. The Disk Chopper Spectrometer (DCS) \cite{copley2003disk} at the NCNR was used to obtain low energy inelastic measurements in the [HHL] plane. Powder averaging over a wide [HHH] direction within the [HHL] plane, the data shows a nearly dispersionless gap (Fig~\ref{fig9}). While this feature should contain the spin ice structure factor, the statistics were too poor to resolve the Q dependence completely within the scattering plane.  However this scattering is almost certainly associated with the energy-integrated scattering observed with DNS.  Integration over [HH0], [HHH] and [00L] high symmetry directions confirm that the scattering is slightly anisotropic, and consistent with spin ice scattering (Appendix). To obtain the magnitude of the inelastic scattering the single Q inelastic scattering normalization method presented by G. Xu (2013) \cite{xu2013absolute} was used over a Q-range of 0.4-0.8 \AA$^{-1}$, using Gaussian fitting for both the elastic and inelastic excitations. This yields an inelastic moment of 0.26(2)~$\mu_B$, which is consistent with a much larger component of the moment remaining static compared to other Nd$^{+3}$ pyrochlores. The magnon spectrum seen in other neodymium pyrochlore systems was not observed here, potentially due to the chemical disorder preventing the long-range propagation of magnons below T$_N$. 

\subsection{Moment Fragmentation}

The formalism of Benton, the ground state exhibited is a direct result of m$_{j} = \{\frac{9}{2},\frac{3}{2}\}$ within the $\overline{D_{3d}}$ double group \cite{benton2016quantum}.   Naively, one would expect that breaking this local symmetry, as in the case of local chemical disorder that is seen in Nd$_2$ScNbO$_7$, should result in a fully ordered Ising antiferromagnet, or a magnetic glass ground state, due to the removal of the symmetry constraints that give rise to dipole-octupole interactions in Nd$^{+3}$ pyrochlores \cite{benton2016quantum}.    However, it is found in this work that this is not the case.   Some of the Nd$^{+3}$ ions still show the moment fragmentation-like behaviour at low temperatures despite chemical disorder which reduces the symmetry to $D_3$ (which is confirmed by the CEF measurements presented here).   This involves an inversion symmetry operation being removed for many of the local Nd sites.   As far as we are aware there have not been any reports in the literature discussing dipole-octupole magnetic states under this reduced $D_3$ symmetry. However, from the symmetry argument presented by Huang \cite{huang2014quantum} it appears that the removal of inversion symmetry should maintain the dipole-octupole symmetry required for the fragmentation-like behaviour observed in other Nd$^{+3}$ systems.   The argument for this is as follows:  the fundamental symmetry requirement of the moment fragmentation-like behaviour is that the x and z components of the moment transform like magnetic dipoles and the y component transforms like a magnetic octupole.  This is the case under D$_{3d}$ symmetry with the ground state crystal field doublet having a m$_j$ of 3n/2 where n is an odd integer, which yields a doublet of $\Gamma_{5}^{+}$ and $\Gamma_{6}^{+}$.   In Nd$_2$ScNbO$_7$, the local B environments that retain D$_{3d}$ symmetry are very rare.  Instead a local environment of alternating Sc and Nb ions are proposed as retaining the dipole-octupole symmetry. This causes a symmetry reduction to D$_3$, removing inversion symmetry.  However under this symmetry reduction $\Gamma_{5}^{+} \Rightarrow \Gamma_{5}$ and $\Gamma_{6}^{+} \Rightarrow \Gamma_{6}$ retain the same symmetry, excluding inversion, which still yields the dipole-octupole symmetry for Benton's description of these Nd$^{+3}$ systems.  Under these conditions the dipole-octupole doublet transforms like

%

\begin{eqnarray}
C_3: \tau^{x,y,z} \Rightarrow \tau^{x,y,z} \\
\sigma_d: \tau^{x,z} \Rightarrow -\tau^{x,z} ~~~ \tau^{y} \Rightarrow \tau^y \\
I: \tau^{x,y,z} \Rightarrow \tau^{x,y,z} \\
\end{eqnarray}
from Chen \cite{huang2014quantum}, and consequently
\begin{equation}
C_2:  \tau^{x,z} \Rightarrow -\tau^{x,z} \\
~~~ \tau^{y} \Rightarrow \tau^y \\
\end{equation}
as
\begin{equation}
C_2 = I~\otimes\ _{\perp}\sigma_d
\end{equation}

An important consequence of this transformation is that $\tau^x$ and $\tau^z$ transform dipoles and $\tau^y$ an octupole creating the dipole-octupole doublet and allowing for the moment fragmentation-like phenomenon.  Therefore the reduction to D$_3$ symmetry should not prohibit the existence of moment fragmentation.


Following this analysis, Nd$_2$ScNbO$_7$ should nominally exhibit the experimental signatures observed in the Nd$_2$B$_2$O$_7$ parent compounds.  However, local distortions induced by the solid state solution of non-magnetic cations break the symmetry requirements for a dipole-octupole ground state which is evidenced by the fact that 2.2(4) $\mu_{B}$ orders into an all-in, all-out Ising antiferromagnet.  By CEF spectroscopy we can observe that 14(2) $\%$ of Nd ions do in fact retain $D_3$ symmetry.  This small fraction of Nd$^{+3}$ ions are still able to show signatures of the `divergence-free' gapped state with a spin ice structure factor, which can be weakly observed in our polarized and time of flight neutron scattering measurements. This moment agrees reasonably well with the dynamic moment observed in the DCS data (0.26(2) $\mu_B$) suggesting that roughly 10\% of the total moment remains dynamic which is consistent with a small fraction of the Nd$^{+3}$ ions exhibiting moment fragmentation-like behaviour. It is also possible that the Nd$^{3}$ universally experience a smaller octupole exchange contribution than in the parent compound, although this is inconsistent with a disorder-driven effect, which we are assuming is the primary driver of the differences between the reasonably well-ordered parent Nd$_2$B$_2$O$_7$ compounds and the ion-disodered Nd$_2$ScNbO$_7$. 

The issue of structural disorder in zirconium pyrochlores has been investigated by a variety of chemical probes for decades.   For example, the work by Blanchard ${et~ al}$ on the rare earth series of pyrochlores has shown definitively that there is a prominent fluorite to pyrochlore crystal structure change at Tb due to the lanthanide contraction \cite{blanchard2012does}. This site disorder problem has been noted in other pyrochlores with relatively large B site cations, such as the plumbates (RE$_2$Pb$_2$O$_7$, RE = rare earth).   In this case, the Nd and Pr have low amounts of A/B site mixing (on the order of a few percent), but by the Gd member of the plumbate series, the chemical disorder is significant (on the order of 30 percent) \cite{HallasPb}.    However, within all the members of the zirconium series, previous experiments have shown that there is a considerable amount of chemical disorder.  As in the case of the plumbates, this site mixing is minimized for early members of the series such as La, Pr and Nd, but nonetheless local probes such as XANES show that even for these early members in the series, there is a significant amount of A/B site mixing.   This experimental work illustrates that even for maximal B-site disorder, the octupolar interactions can occur, and chemical issues surrounding more well-ordered members of the series (such as Nd$_2$Zr$_2$O$_7$) should not play a significant role in observing this ground state.

\section{Conclusions}

Structural disorder is well known to cause magnetic disorder, such as in spin glasses.  Nd$_2$ScNbO$_7$ presents an interesting case where in the presence of severe structural disorder, a remnant of the spin ice `divergence-free' scattering still exists at low temperatures.  While Nd$_2$ScNbO$_7$ is an extreme example of chemical disorder among the rare earth pyrochlores, this system provides insight into the effect of chemical disorder on disordered magnetic states in general.  Many studies of disorder focus on the missing occupation of magnetic ions from the A-site or addition of magnetic cations to the B-site \cite{ehlers2008dynamic, ross2012lightly, ueland2010coexisting, baroudi2015symmetry}.  Equally important are studies of structural distortions that break or maintain symmetry \cite{trump2018universal}, and magnetic disorder can be robust against large structural perturbations assuming an appropriate symmetry is maintained. Additionally this study provides insight into the symmetry requirements for the dipole-octupole ground state in pyrochlores. The fragmentation-like behaviour may occur in systems of $D_3$ symmetry on the pyrochore lattice, with broken inversion symmetry, and the destruction of the local C$_3$ axis allows the system to order into a static Ising antiferromagnet with the coexistence of diffuse scattering.  Additionally our spectroscopic data suggests that Nd$_2$ScNbO$_7$, and likely other charge disordered pyrochlores have charge ice short-range correlations, adding ionic charge ice to the varied types of systems that exhibit ice-like correlations.

\section{Acknowledgements}

The authors would like to thank Jeffrey G. Rau for his important insights, and the experimental aid provided by Graham King.  This research was funded in part by the National Science and Engineering Research Council of Canada (NSERC), and the Canadian Foundation for Innovation (CFI). C.R. Wiebe would like to thank the Canadian Institution For Advanced Research (CIFAR), the Canada Research Chair program (Tier II), and the Leverhulme Trust. C. Mauws would like to thank NSERC for the Alexander Graham Bell Scholarship. Q.H. and H.Z. thank the support from NSF-DMR-1350002. The work performed in NHMFL was supported by NSF-DMR-1157490 and the State of Florida. A portion of this research used resources at the SNS, a DOE Office of Science User Facility operated by the Oak Ridge National Laboratory. An additional portion of this work used resources at the NCNR in Gaithersburg, USA (operated by the National Institute of Standards and Technology), and at the Heinz Maier-Leibnitz Zentrum in Garching, Germany. Part or all of the research described in this paper was performed at the Canadian Light Source, a national research facility of the University of Saskatchewan, which is supported by the Canada Foundation for Innovation (CFI), the Natural Sciences and Engineering Research Council (NSERC), the National Research Council (NRC), the Canadian Institutes of Health Research (CIHR), the Government of Saskatchewan, and the University of Saskatchewan.

\bibliography{Nd2ScNbO7}

\section{Appendix}

\subsection{\label{sec: 2}
Crystal Field Analysis   
}

\begin{figure}[h]
\linespread{1}
\par
\includegraphics[width=3.3in]{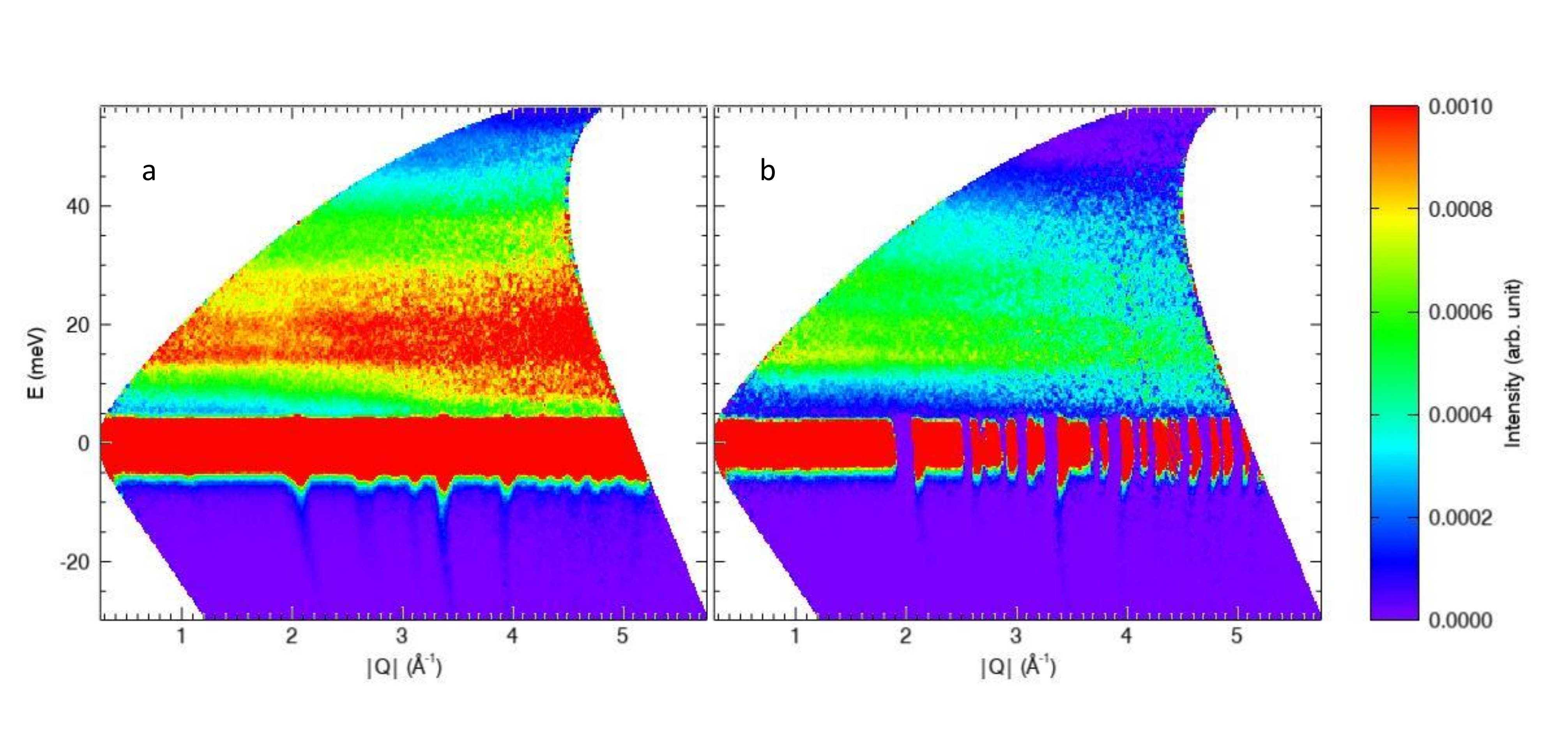}
\par
\caption{Inelastic powder neutron scattering from Nd$_2$ScNbO$_7$ using 60 meV incident energy on the SEQUOIA spectrometer. Raw data (a), and data with a self-shielding factor corrected La$_2$ScNbO$_7$ subtraction to correct for phonon interference. }
\label{SI_CEF}
\end{figure}

\begin{figure}[h]
\linespread{1}
\par
\includegraphics[width=3.3in]{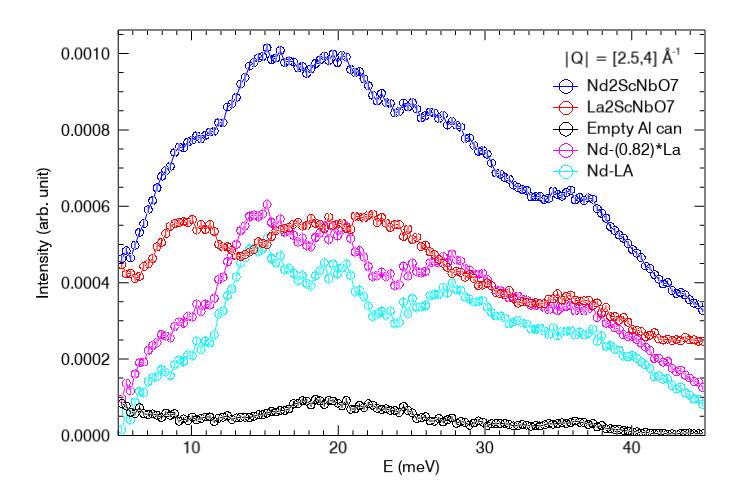}
\par
\caption{Cuts along energy from the 60 meV incident energy data from the SEQUOIA spectrometer, showing the raw data of La$_2$ScNbO$_7$, the direct subtraction and the 0.82 self-shielding factor corrected subtraction.} 
\label{MultiECut}
\end{figure}

\begin{figure}[h]
\linespread{1}
\par
\includegraphics[width=3.3in]{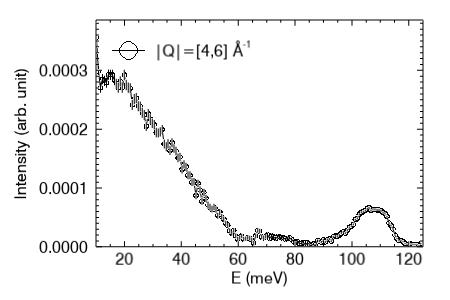}
\par
\caption{Inelastic powder neutron scattering from Nd$_2$ScNbO$_7$ using 140 meV incident energy from the SEQUOIA spectrometer, after phonon subtraction
} 
\label{140meV}
\end{figure}

As stated in the primary text, the majority of the crystal field spectral weight  is distributed continuously from 10-50 meV. To accurately remove the phonon interference we subtract off a La$_{2}$ScNbO$_{7}$ phonon standard, normalized by calculated Bragg peak intensities to a self shielding factor of 0.82 (Fig. \ref{SI_CEF}). The remaining scattering is solely due to crystal field excitations. Additionally the higher energy excitation present at 106.5(1) meV is shown (Fig. \ref{140meV}, and remains very broad at 14.1(2) meV FWHM. At an incident energy of 140 meV the sharp excitations cannot be resolved from the broad features, and the lower broad excitation is not resolved from the elastic line, which is why the relative intensity of the 106.5 meV peak is not fit. 

\begin{figure}[h]
\linespread{1}
\par
\includegraphics[width=3.3in]{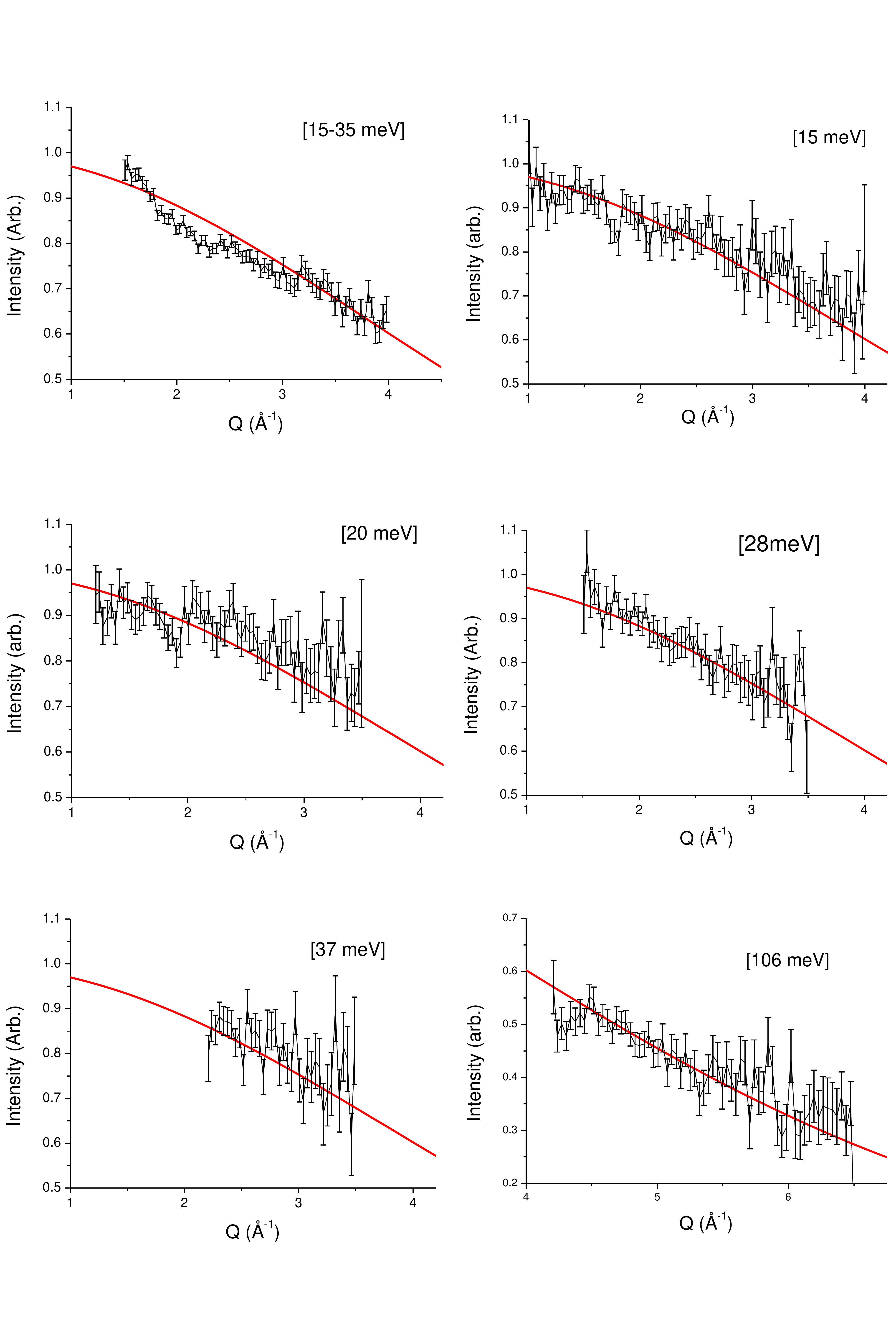}
\par
\caption{Q-dependent cuts of the various inelastic features (black data) from SEQUOIA, with the squared magnetic form factor of Nd$^{+3}$ overlayed. unless otherwise mentioned, enegy integration ranges are $\pm$1 meV.} 
\label{NdFF}
\end{figure}

\begin{figure}[h]
\includegraphics[width=3.3in]{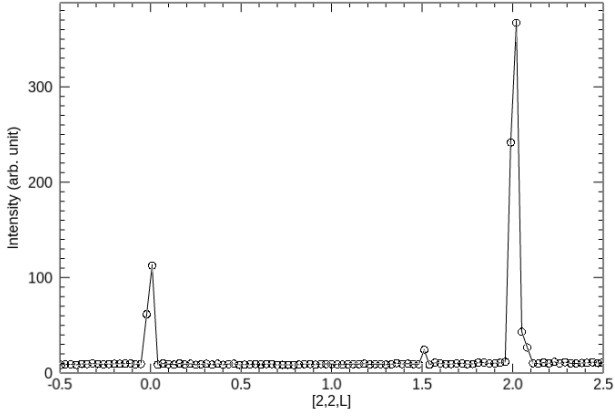}
\caption{Cuts of the Bragg peaks from DCS at 60~mK, including the (220) peak and (222) peak. An incident energy of 5 \AA was used, the cut was taken from the elastic line with an integration of $\pm$100 $\mu$eV, and $\pm$ 0.1 r.l.u. in [HH0]. }
\label{dcscut}
\end{figure}

\begin{figure}[t]
\includegraphics[width=3.3in]{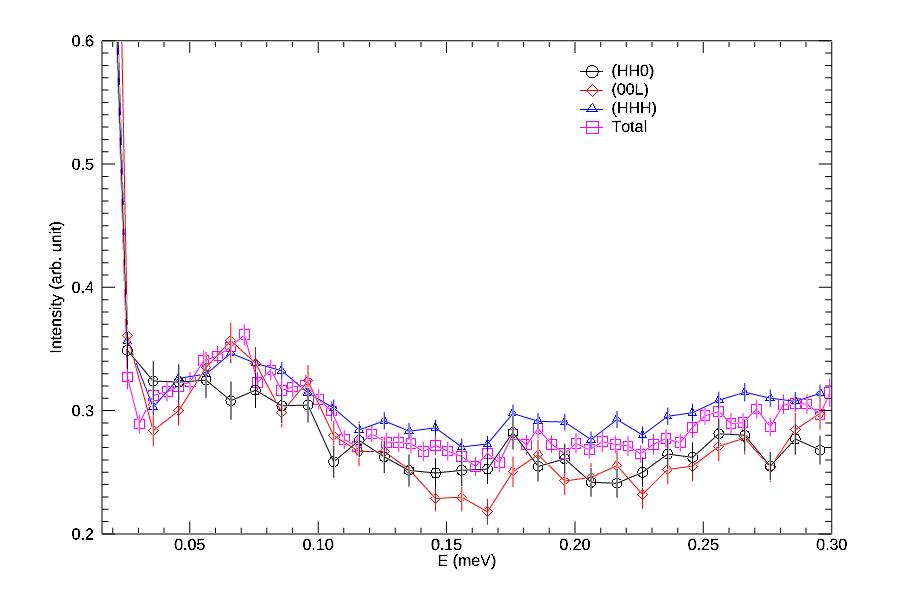}
\par
\caption{Inelastic neutron data taken from the DCS, using an incident wavelength of 8 \AA. Cuts are taken across a variety of directions in the HHL planes, using a $\pm 0.25$ r.l.u. integration in the orthogonal direction.}
\label{dcsvar}
\end{figure}

The Q-dependance of the remaining scattering broad feature scattering (15-35 meV integration window) can be seen to follow the Nd$^{+3}$ form factor (Fig. \ref{NdFF}). Additionally our observed crystal field excitations also follow this behaviour. The peak at 37 meV on the other hand has been assigned as an imperfectly subtracted phonon, as it also appears as a peak in La$_2$ScNbO$_7$. Although it appears to follow the form factor reasonably well, this is likely because it is sitting on the broad excitation. Additionally the phonon appears to be shifted lower in energy when compared to La$_2$ScNbO$_7$ suggesting the improper subtraction is due to a shift in energy, not intensity. Therefore integration over a wide window will not show a phonon-like Q-dependence.  

An additional note on the statistical charge ice model presented in the paper; The expected distribution of 17.6 \% of Nd ions containing the D$_3$ symmetry comes from simply assuming that each Sc ion has a 2/3 chance of an adjacent ion being Nb and vice versa, which is true within charge ice. In order to replicate our value of 14 \% of Nd ions containing the D$_3$ local environment a Monte Carlo simulation was used using the 6 local B-site tetrahedra that surround the A-site. By introducing defects into this model we approach our experimental value. Assuming the interpretation of the system as a charge-ice with defects is correct, then the defect concentration is 0.15 defects per B-site tetrahedron (where a tetrahedron of all Nb or all Sc counts as two defects). Defect configurations were allowed into the simulation with a variable probability that falls off exponentially with the number of defects present.  The probability was tuned to reproduce 14 \% of Nd ions containing the D$_3$ configuration. 

\subsection{\label{sec: 3}
Elastic scattering
}

The Bragg peaks observed in the diffractions data from the DCS and DNS show resolution limited scattering (Fig. \ref{dcscut}), establishing that the elastic peaks are indicative of static, long-range order. DCS data are depicted as they show much higher resolution compared to the DNS.

\subsection{\label{sec: 4}
Anisotropy of Gapped state
}
{

The excitations being seen in DCS are the same signal that gives the spin-ice like scattering from the polarized diffraction data, as this is consistent with moment fragmentation. The signal from DCS was too low to resolve constant energy cuts, however using wide integration windows we can observe directional anisotropy that is consistent with spin ice (Fig. \ref{dcsvar}). Increased scattering along the [H,H,H] and [H,H,0] compared to [0,0,L] is consistent with moment fragmentation and the polarized diffuse scattering data, demonstrating that this feature was in fact inelastic. 

As mentioned in the primary text, the statistics from the DCS were too poor to resolve the structure factor of the gapped excitations. Using large integration windows the excitation shows the correct anisotropy for a spin ice with increased scattering along the [H,H,H] and [0,0,L] directions.


\end{document}